\newlength{\extralineskip}
\documentclass[12pt]{article}

\begin{document}
\begin{titlepage}
\begin{flushright}
          \begin{minipage}[t]{12em}
          \large UAB--FT--421\\
                 July 1997
          \end{minipage}
\end{flushright}
\vspace{\fill}

\vspace{\fill}

\begin{center}
\baselineskip=2.5em

{\large \bf Gravitinos from Gravitational
Collapse}
\end{center}

\vspace{\fill}

\begin{center}
{\bf J.A. Grifols, E. Mass\'o, and R. Toldr\`a}\\
\vspace{0.4cm}
     {\em Grup de F\'\i sica Te\`orica and Institut de F\'\i sica
     d'Altes Energies\\
     Universitat Aut\`onoma de Barcelona\\
     08193 Bellaterra, Barcelona, Spain}
\end{center}
\vspace{\fill}

\begin{center}
\large Abstract
\end{center}
\begin{center}
\begin{minipage}[t]{36em}
We reanalyse the limits on the gravitino mass $m_{3/2}$ in superlight gravitino
scenarios derived from arguments on energy-loss during gravitational collapse. We conclude
that the mass range $10^{-6}eV\leq m_{3/2}\leq2.3\times10^{-5}eV$ is excluded by SN1987A
data. In terms of the scale of supersymmetry breaking $\Lambda$, the range
$70GeV\leq\Lambda \leq 300GeV$ is not allowed.
\end{minipage}
\end{center}

\vspace{\fill}

\end{titlepage}

\clearpage

\addtolength{\baselineskip}{\extralineskip}

In a wide class of supergravity models with SUSY breaking $\Lambda$  in the TeV range,
the gravitino can be very light:
\begin{equation} \label{mass}
m_{3/2}=2.5 \times 10^{-4} eV    (\Lambda/1 TeV)^2.
\end{equation}
Indeed, models where gauge interactions mediate the breakdown of supersymmetry
\cite{dine}, models where an anomalous U(1) gauge symmetry induces SUSY breaking
\cite{bine}, and no-scale models are all examples of models where a superlight gravitino
can be accommodated \cite{ellis}. In all of them, the gravitino is the LSP and,
furthermore, its couplings to matter and radiation are inversely proportional to its
mass. Therefore, one may expect interesting phenomenology \cite{kim}.
 Bounds on the gravitino mass, or equivalently on the scale $\Lambda$ have been given in
the context of those models by various authors and have been extracted from different
physical systems. In fact, the limits come from as distinct areas as the anomalous
magnetic moment of the muon \cite{men}, primordial nucleosynthesis \cite{fayet} or
stellar energy drain arguments \cite{nowa}.
 In recent papers \cite{luty}, it has been noted that the amplitudes for gravitino
processes that were used in deriving some of the constraints had an incorrect energy
behaviour. In particular, the supernova (SN) bounds deduced in ref.\cite{moha} using 
the effective couplings given explicitly by Gherghetta \cite{gergy} are invalid as
pointed out by Luty and Ponton \cite{lute}. These authors, however, when reexamining the
limits coming from the SN1987A explosion, use an incorrect abundance of positrons in the
core, do not discuss gravitino bremsstrahlung, and misidentify the main
source of opacity in the stellar core. The purpose of the present paper is thus to redo
the analysis that renders the bounds on $m_{3/2}$(or $\Lambda$) following from SN
collapse. Since SN considerations gave the best limits on $\Lambda$ up to now \cite{lute},
this is not an empty exercise.
  
 The relevant piece in the effective lagrangean is the derivative coupling of the
goldstino $\chi$  to photons:
\begin{equation} \label{coupl}
\delta L_{eff} =(e/2)(M/\Lambda^2)^2\partial^\mu\chi\sigma^\nu \bar{\chi} F_{\mu\nu}+h.c.
\end{equation}
with $F_{\mu\nu}$, the electromagnetic field-strength and $M$ is a mass that depends on
the supersymmetry breaking model. In gauge-mediated models, $M \sim m_{\tilde L}/4\pi$,
where $m_{\tilde L}$ is the left-handed slepton mass.
 Given that gravitino pairs are mainly produced via one-photon interactions, the sources
of gravitino luminosity in stars are, in principle, gravitino bremsstrahlung in
neutron-proton scattering, pair production in electron-positron annihilation and plasmon
decay into gravitinos.
 The energy-loss rate (per unit volume) via $ pn\rightarrow pn \tilde{G}\tilde{G} $ is,
\begin{equation} \label{brems}
Q=\int  {d^3k_1\over(2\pi)^3 2k_1^0}{d^3k_2\over(2\pi)^3
2k_2^0}\prod_{i=1}^4{d^3p_i\over(2\pi)^3 2p_i^0}f_1f_2(1-f_3)(1-f_4)(2\pi)^4
\delta^4(P_f-P_i)\sum_{spins} \mid M_{fi} \mid^2(k_1^0+k_2^0)
\end{equation}
where $(p^0,\vec{p})_i$ are the 4-momenta of the initial and final state nucleons,
$(k^0,\vec{k})_{1,2}$ are the 4-momenta of the gravitinos and $ f_{1,2}$ are the
Fermi-Dirac distribution functions for the initial proton and neutron and $(1-f_{3,4})$
are the final Pauli blocking factors for the final proton and neutron. The squared matrix
element can be factorised as follows,
\begin{equation} \label{factor}
\sum_{spins} \mid M_{fi} \mid ^2=(2\pi)^2\alpha^2(M/\Lambda^2)^4N_{\mu\nu}
G_{3/2}^{\mu\nu}
\end{equation}
where $N_{\mu\nu}$ is the nuclear (OPE) tensor and $G_{3/2}^{\mu\nu}$ is the gravitino
tensor in the matrix element squared. The factor $N_{\mu\nu}$ is common to any
bremsstrahlung process involving nucleons.It appears, e.g., in neutrino bremsstrahlung
calculations and in axion bremsstrahlung calculations, and is given explicitly in ref
\cite{raffelt}.On the other hand, $G_{3/2}^{\mu\nu}$ is a tensor specific to gravitino
bremsstrahlung. It reads,
\begin{equation} \label{gravten}
G_{3/2}^{\mu\nu}=k_{1}^\mu k_{2}^\nu+k_{2}^\mu k_{1}^\nu-k_{1}.k_{2}g^{\mu\nu}
\end{equation}
The integration of $N_{\mu\nu}$ over the phase-space of the nucleons can be performed
explicitly and the details can be found again in Raffelt's book \cite{raffelt}. When we
contract the result with the gravitino tensor $G_{3/2}$ and perform the integrals over
gravitino momenta to complete the energy depletion rate, we are led to the following
emissivity:
\begin{equation} \label{emiun}
Q_{brems}^{ND}
=(8192/385\pi^{3/2})\alpha^2\alpha_{\pi}^2(M/\Lambda^2)^4Y_{e}n^2_BT^{11/2}/m^{5/2}_p
\end{equation}
for non-degenerate and non-relativistic nucleons ($\alpha_\pi$ is the pionic
fine-structure constant, $n_B$ is the number density of baryons, and $Y_e$ is the mass
fraction of protons). However, nucleons are moderately degenerate in the SN core. The
emissivity in the (extreme) degenerate case is calculated to be,
\begin{equation} \label{emidos}
Q_{brems}^D=(164\pi^3/4725)\alpha^2\alpha_\pi^2(M/\Lambda^2)^4p_FT^8
\end{equation}
with $p_F$, the Fermi momentum of the nucleons. Numerically, for the actual conditions of
the star, both emissivities differ by less than an order of magnitude (about a factor of
three). Since the actual emissivity interpolates between these two values, we shall adopt
the smallest of the two (i.e. $Q^{ND}_{brems}$) to make our (conservative) estimates. We
turn next to the annihilation process.

 The energy loss for the process $e^+(p_1)+e^-(p_2) \rightarrow
\tilde{G}(k_1)+\tilde{G}(k_2)$ can be calculated along similar lines as above. The spin
averaged matrix element squared is in this case,
\begin{equation} \label{matrix}
\sum_{spins}\mid
M_{fi}\mid^2=(2\pi)^2\alpha^2(M/\Lambda^2)^4E_{\mu\nu}(p_1,p_2)G_{3/2}^{\mu\nu}(k_1,k_2)
\end{equation}
where $E_{\mu\nu}(p_1,p_2)$ equals formally the tensor $G_{3/2}^{\mu\nu}$ in eq.(5) with
$k_1,k_2$ replaced by $p_1,p_2$.
The luminosity then is found to be,
\begin{equation} \label{lum}
Q_{ann}=8\alpha^2(M/\Lambda^2)^4T^4e^{-\mu/T}\mu^5b(\mu/T)/15\pi^3
\end{equation}
with $b(y)\equiv(5/6)e^yy^{-5}(F_5^+ F_4^- +F_4^+F_5^-)$
where 
$F_{m}^\pm(y)=\int_0^\infty dxx^{m-1}/(1+e^{x\pm y})$ ($\mu$ is the chemical
potential of the electrons). The function
$b(y)\rightarrow1$ in the degenerate limit.
 Finally, our estimate of the plasmon decay luminosity is,
\begin{equation} \label{plasm}
Q_P=16\zeta(3)\alpha^4T^3\mu^6(M/\Lambda^2)^4/81\pi^5
\end{equation}
(where only transverse plasmons have been taken into account).

Taken at face value, the bremsstrahlung rate is the largest of the three. However,
$Q_{brems}$ is overestimated since we did not consider multiple scattering
effects which are present in a dense medium\cite{raffelt}. Indeed, as for the axion
case\cite{raffelt},the gravitino bremsstrahlung rate probably saturates around $20\%$
nuclear density and this should be taken into account when evaluating eq.(6). If we use
now the values
$T=50 MeV$,
$\mu=300 MeV$, and
$Y_e=0.3$, eqs. (6) (with $n_B\sim 0.2 n_{nuc}$), (9) and (10) give 
\begin{equation} \label{ratios}
Q_{ann}:Q_{brems}:Q_P\approx   1.2\times10^3:3\times10^2:1
\end{equation}

Therefore, a limit on $\Lambda$
will follow from the requirement that $L_{3/2} \approx VQ_{ann}$ ($V$ is the volume of
the stellar core) should not exceed $10^{52}ergs/s$. This constraint on the gravitino
luminosity
$L_{3/2}$ implies, in turn,
\begin{equation} \label{bound}
\Lambda\geq300 GeV (M/43 GeV)^{1/2}(T/50 MeV)^{11/16}(R_c/10 Km)^{3/8}
\end{equation}
or, using eq.(\ref{mass}),
\begin{equation} \label{mass2}
m_{3/2}\geq2.3\times10^{-5}eV.
\end{equation}

Of course, the previous calculation makes sense only if gravitinos, once produced,
stream freely out of the star without rescattering. That they actually do so, for
$\Lambda\geq300 GeV$, can be easily checked by considering their mean-free-path in the
core. The main source of opacity for gravitinos is the elastic scattering off the Coulomb
field of the protons:
\begin{equation} \label{mfp}
\lambda=1/\sigma n=(4/\pi\alpha^2)Y_e^{-1}\rho^{-1}m_p^{-1}(\Lambda^2/M)^4
\end{equation}
 The thermally averaged cross-section for elastic gravitino scattering on electrons is
roughly a factor $T\mu/m_p^2$ smaller than that on protons and thus it does not
contribute appreciably to the opacity.
 Putting numbers in eq.(14) we find:
\begin{equation} \label{mfp2}
\lambda \simeq1.4\times10^7 cm(43 GeV/M)^4(\Lambda/300 GeV)^8
\end{equation}

 On the other hand, the calculation of Q breaks down for $\lambda\leq10Km$, i.e. for
$\Lambda\leq220GeV$, when gravitinos are trapped in the SN core. In this case,
gravitinos diffuse out of the dense stellar interior and are thermally radiated from a
gravitino-sphere $R_{3/2}$. Because in this instance the luminosity is proportional to
$T^4$, only for a sufficiently large $R_{3/2}$(where the temperature is correspondingly
lower), the emitted power will fall again below the nominal $10^{52}erg/s$. Consequently,
gravitino emission will be energetically possible, if $\Lambda$ is small enough.
 The gravitino-sphere radius can be computed from the requirement that the optical depth
\begin{equation} \label{depth}
\tau=\int_R^\infty dr/\lambda(r)
\end{equation}
be  equal to $2/3$ at $R=R_{3/2}$. Here, $\lambda(r)$ is given in eq.(14) with the density
profile ansatz:
\begin{equation} \label{rho}
\rho(r)=\rho_c(R_c/r)^m
\end{equation}
with $\rho_c=8\times10^{14}g/cm^3$, $R_c=10Km$ and $m=5-7$ and which satisfactorily
parameterises the basic properties of SN1987A \cite{turner}.
 An explicit calculation renders:
\begin{equation} \label{sphere}
R_{3/2}=R_c[(8Y_e/3\pi\alpha^2)(\Lambda^2/M)^4(m-1)/\rho_cR_cm_p]^{1/1-m}
\end{equation}
 Stefan-Boltzmann's law implies for the ratio of gravitino to neutrino luminosities,
\begin{equation} \label{SB}
L_{3/2}/L_\nu=(R_{3/2}/R_\nu)^2[T(R_{3/2})/T(R_\nu)]^4
\end{equation}
where $R_\nu$ is the radius of the neutrinosphere. To proceed further we use the
temperature profile:
\begin{equation} \label{temp}
T=T_c(R_c/r)^{m/3}
\end{equation}
which is a consequence of eq.(17) and the assumption of local thermal equilibrium. Now,
taking $m=7$\cite {mogri}, we obtain
\begin{equation} \label{SB2}
L_{3/2}/L_\nu=(R_\nu/R_c)^{22/3}[(16Y_e/\pi\alpha^2)(\Lambda^2/M)^4/\rho_cR_cm_p]^{11/9}
\end{equation}
 By demanding that $L_{3/2}\leq0.1L_\nu$ and using $R_\nu\simeq30Km$, we get
\begin{equation} \label{upbnd}
\Lambda\leq70GeV.
\end{equation}
This in turn implies $m_{3/2}\leq10^{-6}eV$. Since, on the other hand, the anomalous
magnetic moment of the muon already requires $m_{3/2}$ to be larger than
$\sim10^{-6}eV$ \cite{men,ferrer}, we are forced to conclude that 
\begin{equation} \label{lobnd}
\Lambda\geq300GeV
\end{equation}
or, equivalently,
\begin{equation} \label{last}
m_{3/2}\geq2.3\times10^{-5}eV.
\end{equation}

 In conclusion, we have carefully rederived the bounds on the superlight gravitino
mass (i.e. the SUSY scale $\Lambda$) that follow from SN physics. These limits are
completely general in the sense that they do not rely on other particles in a given
particular model  being light. Should other particles such as the scalar partners of the
goldstino also be light, then the resulting bounds are necessarily tighter. In such
clearly less general frame, constraints have also been derived in the literature
\cite{grif} that are not subject to the criticisms mentioned in the beginning of this
paper. They are much stronger then the ones given here and typically give $\Lambda\geq
300TeV$ (or,$m_{3/2}\geq 50eV$ ) from stellar (e.g. the Sun) evolution arguments,
provided
$m_{3/2}\leq 1KeV$ (e.g. $T_\odot$).

Work partially supported by the CICYT Research Projects AEN95-0815 and AEN95-0882 and the Theoretical
Astroparticle Network under the  EEC Contract No. CHRX-CT93-0120 (Direction Generale 12 COMA).

\end{document}